\documentclass[twocolumn,aps,floatfix,superscriptaddress,preprintnumbers,amsmath,amssymb,showpacs,showkeys]{revtex4-1}

\usepackage{graphicx}
\usepackage{float}
\usepackage{tabularx}

\begin{document}

\title{Universal behavior of the IMS domain formation in superconducting niobium}

\author{A. Backs}
\email{alexander.Backs@frm2.tum.de}
\affiliation{Heinz Maier-Leibnitz Zentrum (MLZ), Technische Universit\"at M\"unchen, Lichtenbergstr. 1, Garching, Germany}%
\affiliation{Physik-Department E21, Technische Universit\"at M\"unchen, James-Franck-Str. 1, Garching, Germany}%

\author{M. Schulz}
\affiliation{Heinz Maier-Leibnitz Zentrum (MLZ), Technische Universit\"at M\"unchen, Lichtenbergstr. 1, Garching, Germany}%
\affiliation{Physik-Department E21, Technische Universit\"at M\"unchen, James-Franck-Str. 1, Garching, Germany}%

\author{V. Pipich}
\affiliation{J\"ulich Center for Neutron Science (JCNS), J\"ulich, Germany}

\author{M. Kleinhans}
\affiliation{Physik-Department E51, Technische Universit\"at M\"unchen, James-Franck-Str. 1, Garching, Germany}

\author{P. B\"oni}
\affiliation{Physik-Department E21, Technische Universit\"at M\"unchen, James-Franck-Str. 1, Garching, Germany}

\author{S. M\"uhlbauer}
\email{sebastian.muehlbauer@frm2.tum.de}
\affiliation{Heinz Maier-Leibnitz Zentrum (MLZ), Technische Universit\"at M\"unchen, Lichtenbergstr. 1, Garching, Germany}%

\date{\today}

\begin{abstract}

In the intermediate mixed state (IMS) of type-II/1 superconductors, vortex lattice (VL) and Meissner state domains coexist due to a partially attractive vortex interaction. Using a neutron-based multiscale approach combined with magnetization measurements, we study the continuous decomposition of a homogeneous VL into increasingly dense domains in the IMS in bulk niobium samples of varying purity. We find a universal temperature dependence of the vortex spacing, closely related to the London penetration depth and independent of the external magnetic field. The rearrangement of vortices occurs even in the presence of a flux freezing transition, i.e. pronounced pinning, indicating a breakdown of pinning at the onset of the vortex attraction.

\end{abstract}

\maketitle

% ------------------------------------------------------------------------------
% Introduction
% ------------------------------------------------------------------------------

Conventional superconductors are divided by the Ginzburg-Landau parameter $\kappa$ into type-I ($\kappa< 1/\sqrt{2}$) and type-II ($\kappa> 1/\sqrt{2}$), which, additionally to the Meissner state (MS) exhibit the Shubnikov state (SS). In the SS, magnetic vortices form a variety of vortex matter (VM), such as the Abrikosov vortex lattice (VL)~\cite{1957Abrikosov}, glassy~\cite{1993Cubitt, 2001Klein, 2018ToftPetersen} or liquid~\cite{1988Gammel, 1993VanDerBeek, 1989Fisher} states. Type-II superconductors are further subdivided, where type-II/2 ($\kappa\gg 1/\sqrt{2}$) features a purely repulsive inter-vortex interaction. In type-II/1($\kappa\approx 1/\sqrt{2}$) the interaction acquires an attractive component~\cite{1971Kramer, 1973Auer, 1987Klein}, which favors the formation of vortex clusters, leaving behind MS regions. The resulting domain structure is denoted the intermediate mixed state (IMS). 

The IMS in the type-II/1 superconductor niobium (Nb) is an ongoing research topic since its first observation via Bitter decoration in 1967 \cite{1967Traeuble,1968Sarma}. Despite numerous experimental~\cite{1971Aston, 1973Auer, 1977Christen, 1989Weber, 2003Aegerter} and theoretical~\cite{1971Kramer, 1976Brandt, 1987Klein, 2016Vagov} efforts, its properties are not yet fully understood. The interplay of repulsive and attractive vortex interactions and the consequences on the domain structure in superconducting VM have found renewed interest with the discovery of multiband superconductors, especially MgB$_2$ (sometimes denoted as type-1.5)~\cite{2009Moshchalkov}. Apart from superconducting properties, the IMS is also a model system for universal domain physics~\cite{1995Seul}, as it can be tuned readily by temperature and magnetic field~\cite{2017Reimann}. 

\begin{figure}[h]
	\includegraphics[width=8cm]{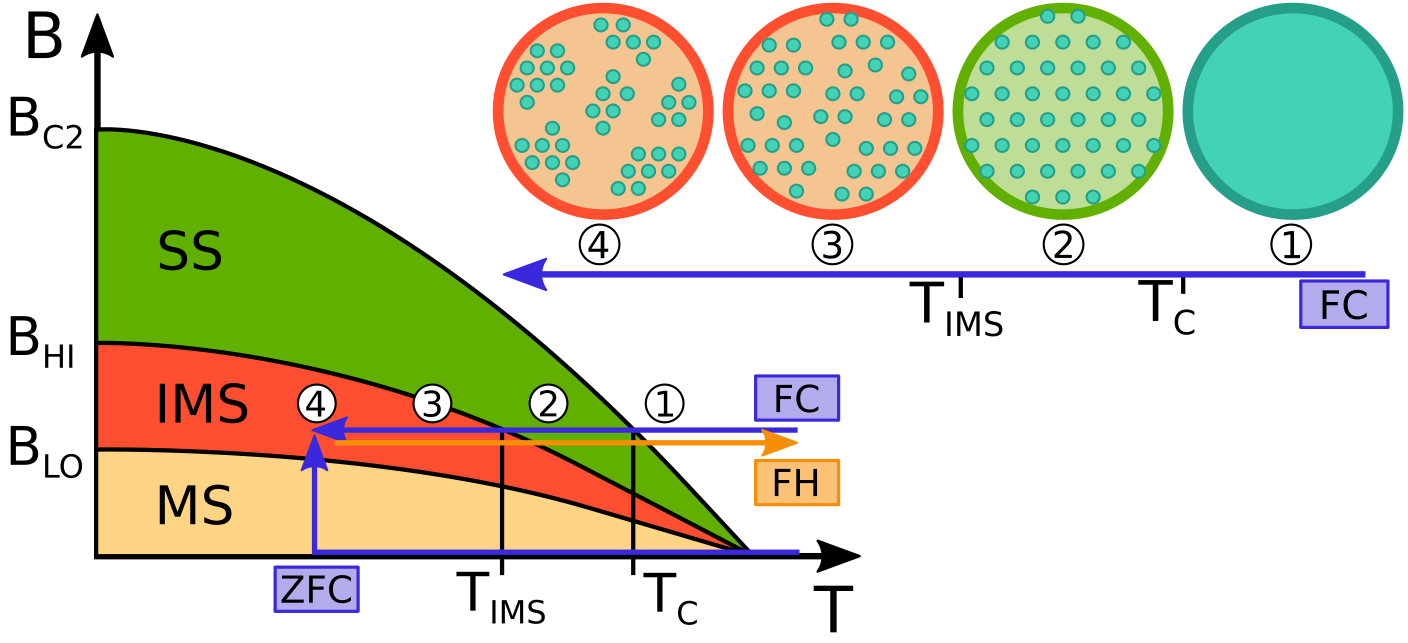}
	\caption{\label{fig:image1} Schematic phase diagram of a type-II/1 superconductor, subdivided into MS, IMS and SS. Arrows depict different measurement protocols: FC, FC/FH and ZFC/FH. For FC measurements, the microscopic magnetic flux redistribution is shown, starting from the normal state (1) with a homogeneous distribution, to the regular VL in the SH (2). In the IMS (3, 4) the VL breaks up into domains containing an increasingly dense VL.}
\end{figure}

Previous studies primarily investigated the zero field cooled (ZFC) field dependence of the IMS. However, this approach leads to strong magnetic inhomogeneities due to geometric and demagnetization effects reflected in the critical state model~\cite{1999Brandt, 2001Brandt}. In contrast, our systematic study focuses on the temperature dependence during a field cooling (FC) and subsequent field heating (FC/FH) protocol in bulk Nb samples with distinct pinning properties. The phase diagram and the transition from a homogeneous VL in the SS to the increasingly dense VL domains in the IMS is sketched in Fig.~\ref{fig:image1} on a FC path.

%******************************************
%******************************************

Using small-angle neutron scattering (SANS), very-small-angle neutron scattering (VSANS) and neutron grating inteferometry (NGI) we cover length scales from $10\,$nm to $10\,\mu$m. Combined with bulk magnetization measurements, we find that in the IMS the vortex attraction leads to a preferred vortex spacing $a_{VL}^{IMS}$, \textit{independent} of the external magnetic field. $a_{VL}^{IMS}$ shows a universal temperature dependence, which is closely related to the superconducting penetration depth $\lambda_L$ and therefore to the sample purity. Remarkably, we find the SS to IMS transition even for samples with pronounced pinning. The continuous microscopic rearrangement of vortices at temperatures \textit{below} a vortex freezing transition is consistent with a putative breakdown of pinning at the onset of the vortex attraction. Accordingly, the \textit{reversibility} of the IMS transition for FC/FH disproved metastability as origin of the observed phenomenon.

% ------------------------------------------------------------------------------
% Experimental details                                            
% ------------------------------------------------------------------------------

The properties of all single crystal niobium samples used in this study are summarized in Tab.1 of the Supplemental Material, including their shape, dimension, demagnetization factor and the corresponding experimental techniques. Nb-hp-1 (high purity) is a custom grown crystal with exceptional quality, used previously~\cite{2009Laver, 2009Muehlbauer,2015Reimann}, which was specifically oxidized to decrease surface pinning. Nb-mp (medium purity) is a large polycrystalline disc obtained from Heraeus with grains with a diameter of $\sim10\,$cm. Nb-lp (low purity) was obtained from Matek and previously used~\cite{2017Reimann}. The samples cut from Nb-mp and Nb-lp have been prepared by spark erosion, diamond wire cutting, grinding, polishing and etching in fluoric acid. Different combinations of polishing and etching had no visible impact on the measurements. The different purities were determined by residual resistivity ratio ($RRR$) measurements and by neutron activation analysis~\cite{2015Revay}. Nb-hp exhibits $RRR>10000$ and $20\,$ppm Ta as primary impurity. Nb-mp was specified with $RRR>300$ and $150\,$ppm Ta. For Nb-lp, $RRR\approx100$ with $200\,$ppm Ta and $350\,$ppm W.

The magnetization was measured with a vibrating sample magnetometer (VSM), equipped with a He-flow cryostat. All neutron experiments (see Tab.2 of the Supplemental Material) were performed at the Heinz Meier-Leibnitz Zentrum (MLZ). For all measurements, a closed cycle cryostat has been employed
and the magnetic field was aligned parallel to the neutron beam. SANS data was obtained at the instrument SANS-1~\cite{2015Heinemann}. Collimation length and sample-detector distance were both $20\,$m, with a source aperture diameter of $20\,$mm, a sample aperture of $7\,$mm and a neutron wavelength of $\lambda=12\,\text{\AA}$ with $\Delta\lambda/\lambda=0.06$. VSANS measurements were performed at the toroidal mirror instrument KWS-3 \cite{2015Pipich}. The instrument collimation was determined by the entrance aperture of $1.5\,$mm$\times 1.5\,$mm, the sample-detector distance was set to $10\,$m and $\lambda=12.8\,\text{\AA}$ with $\Delta\lambda/\lambda=0.2$ was used. NGI measurements were performed at the instrument ANTARES~\cite{2009Calzada, 2015Schulz, 2016Reimann}. The instrumental setup, a collimation of $L/D = 250$, $\lambda=4\,\text{\AA}$ with $\Delta\lambda/\lambda=0.1$ and sample detector distance $d=20\,$cm, yield an NGI correlation length of $1.9\,\mu$m~\cite{2014Strobl}.

% ------------------------------------------------------------------------------
% Results                            
% ------------------------------------------------------------------------------

% VSM

\begin{figure}[h]
	\includegraphics[width=8cm]{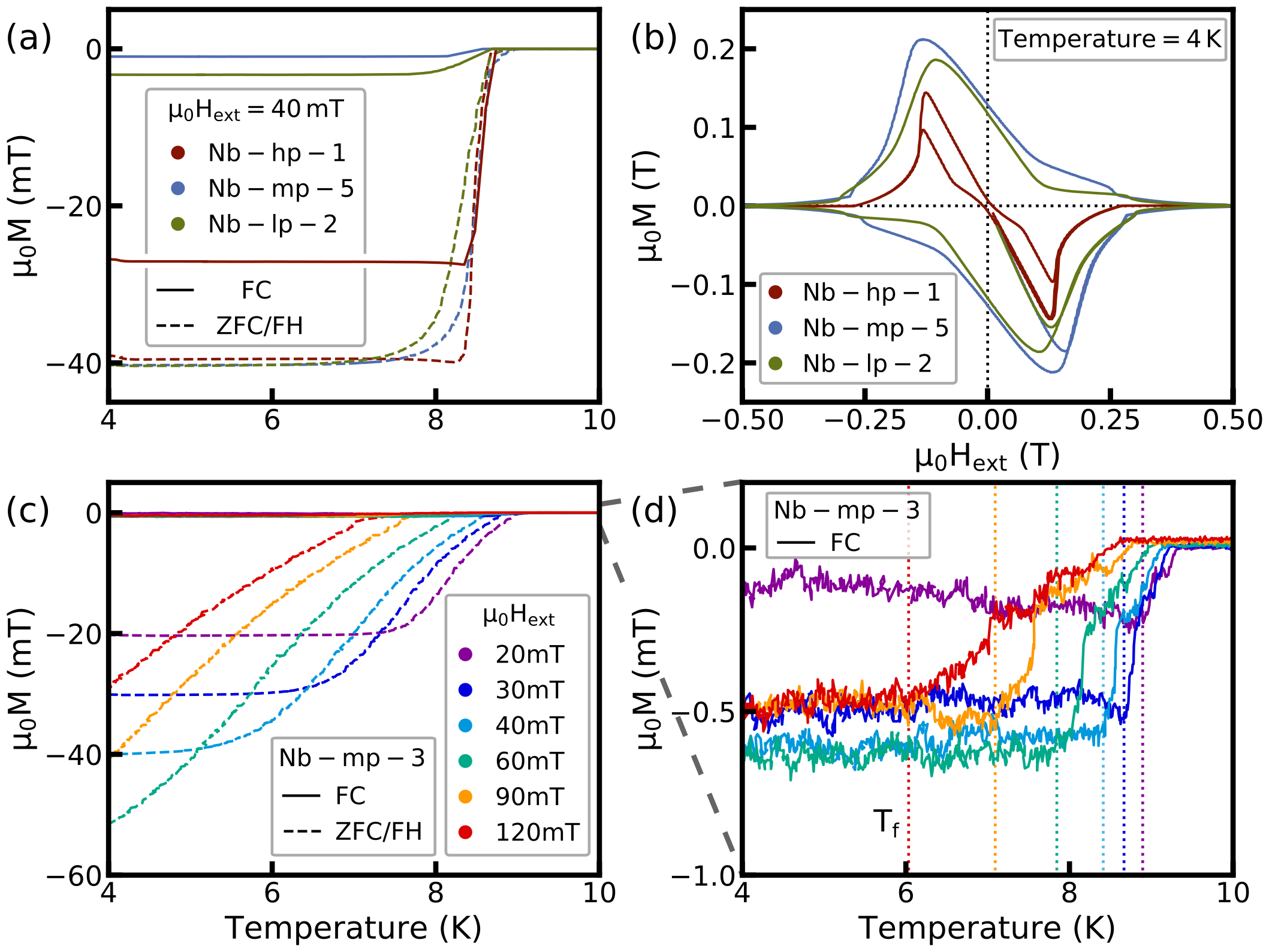}
	\caption{\label{fig:image2} Magnetization data. (a,b) Comparison of the samples Nb-hp-1, Nb-mp-5 and Nb-lp-2. The temperature dependence at $40\,$mT is shown in (a) for FC and ZFC/FH. The field dependence at $4\,$K is shown in (b) in the form of hysteresis loops. (c, d) contain measurements for Nb-mp- 3 at various fields, with FC and ZFC/FH data shown in (c) and a zoom on the field dependent flux freezing transition ($T_f$) in FC measurements in (d).}
\end{figure}

The magnetization measurements shown in Fig.~\ref{fig:image2} illustrate the impact of the sample purity. Field scans (b) are shown as measured, while temperature scans (a, c, d) are corrected for their demagnetization factor ($D$). Panels (a, b) show a comparison between Nb-hp-1, Nb-mp-5 and Nb-lp-2, all oriented with their longest axis parallel to the magnetic field. The temperature dependence is shown in panel (a) for $\mu_0H_{ext}=40\,$mT. At low temperature, FH measurements after ZFC (ZFC/FH, broken lines) start at a constant value of $-\mu_0M=40\,$mT, indicating the perfect diamagnetism of the superconductor. The transition to the normal state, just below $T_c$, depends on the sample quality and is sharpest for Nb-hp and broadest for Nb-lp. For the FC measurements (solid lines) the superconducting transition is at a slightly reduced temperature due to geometric barriers. Below $T_c$, the magnetization decreases, similar to the ZFC/FH data, but saturates at a smaller value. This effect marks the VL freezing transition, where collective pinning traps a part of the magnetic flux inside the material. The temperature of saturation is denoted by $T_f$. The effect of VL freezing is stronger for the lower purity samples (Nb-mp, Nb-lp). Panel (b) shows magnetization loops measured at $4\,$K in the range between $\pm500\,$mT. Nb-mp and Nb-lp show similar broad hysteresis loops, again characteristic of strong pinning, with slight differences due to sample purity and surface quality. In contrast, Nb-hp is perfectly reversible at high fields and only opens up a narrow hysteresis at low fields. 

%******************************************

%******************************************
%******************************************

\begin{figure*}
\includegraphics[width=16cm]{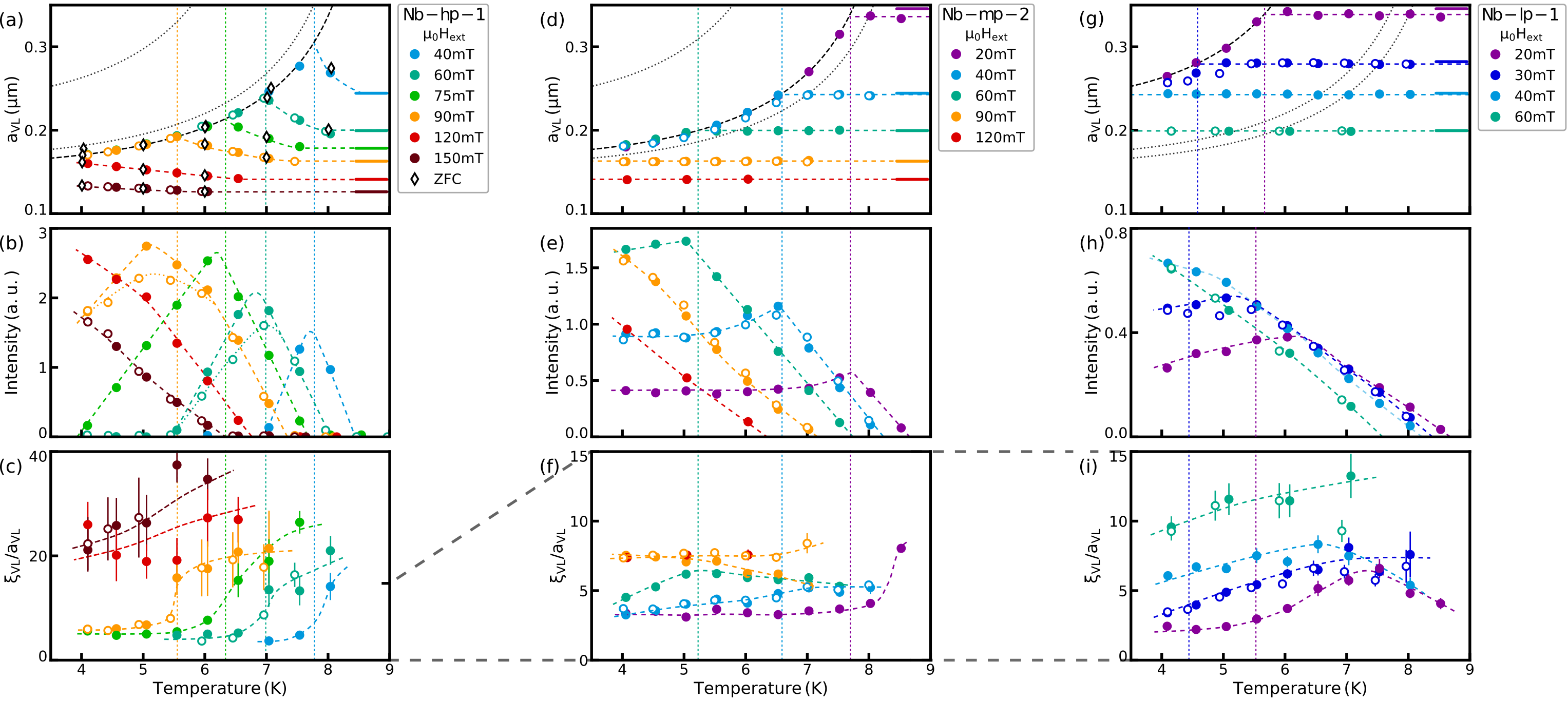}
\caption{\label{fig:image3} Temperature dependent SANS data for Nb-hp-1 (a, b, c), Nb-mp-2 (d, e, f) and Nb-lp-1 (g, h, i). The vortex separation $a_{VL}$ (a, d, g), the integrated scattering intensity (b, e, h) and the VL correlation length $\xi_{VL}$ (c, f, i) were obtained from the first order Bragg peaks. Closed circles are FC measurements, open circles FC/FH and open black diamonds are field scans after ZFC. (a, d, g) include the values of $a_{VL}$ corresponding to $\mu_0H_{ext}$ (thick colored bars) and fits of $a_{VL}^{IMS}$ (black dotted lines). $T_{IMS}$ is indicated by colored vertical dotted lines. Colored lines are guides to the eye.}
\end{figure*}

%******************************************

Panels (c, d) contain detailed measurements on Nb-mp-3. The magnetic field was applied perpendicular to the sample face resulting in a larger $D$. Panel (c) shows the ZFC/FH and FC temperature dependence in magnetic fields up to $120\,$mT. Compared to panel (a), the ZFC/FH measurements have a significantly broadened transition from the superconducting to the normal state, and a strongly enhanced VL freezing transition. Both is in accordance with the critical state model in samples with a large $D$~\cite{1999Brandt, 2001Brandt}. Perfect diamagnetism (i.e. the Meissner state) is observed only below $60\,$mT. A zoom on the FC measurements is shown in panel (d), where the magnetization is very small for all external fields, with $-\mu_0M<1\,$mT. The field dependent $T_f$ is marked for all curves. 

% SANS

The properties of the VL extracted from the SANS data are summarized in Fig.~\ref{fig:image3}. The measurements were performed on Nb-hp-1 with the field perpendicular to the cylinder axis, and Nb-mp-2 and Nb-lp-1 with the field perpendicular to the sample face. Note that demagnetization effects apply only to Nb-hp, as flux freezing prevents demagnetization for Nb-mp and Nb-lp (Fig.~\ref{fig:image2}(d)). SANS data was acquired via rocking scans over $\pm1.5^{\circ}$ with respect to a horizontal and vertical axis, and corrected for a high temperature background ($T=10\,$K). All samples showed a hexagonal pattern of Bragg peaks with angles close to $60^\circ$ and either a single or twofold domains present, as expected from literature~\cite{2009Muehlbauer}. However, in ZFC measurements, a signal indicative of an ordered VL was available only for Nb-hp. From the measurements, the vortex separation $a_{VL}$ was obtained from the Bragg peak positions, in addition to the integrated intensity of the first order peaks and the longitudinal lattice correlation length $\xi_{VL}$. The latter was determined from the Bragg peak FWHM in $\mathbf{q}$-direction and corrected for the instrumental resolution~\cite{1995Yaron} (see Supplemental Material). 

In Fig.~\ref{fig:image3}, full circles correspond to FC data, open circles to FC/FH data and open black diamonds to ZFC data. Colored lines connecting the measurement points are guides to the eye and vertical lines indicate the IMS transition as extracted from $a_{VL}$. At high temperature, all measurements of $a_{VL}$ (panels (a, d, g)) start at values corresponding to the external field (indicated by thick colored bars). Upon cooldown, either an increase due to diamagnetic flux expulsion (Nb-hp) or a constant value in the case of flux freezing (Nb-mp, Nb-lp) is observed. At low external fields, all samples exhibit a downward kink of $a_{VL}$, which is identified with the transition to the IMS ($T_{IMS}$), followed by a continuous decrease to the lowest temperatures. Note that for Nb-mp and Nb-lp, $T_{IMS}(B)<T_f(B)$. In the IMS, $a_{VL}$ is independent of the external field but follows a unique temperature dependence (black dotted lines, see discussion). 

Starting at high temperature, the integrated intensity (panels (b, e, h)) is linearly increasing in the SS, as expected from literature due to a temperature dependent vortex form factor~\cite{1975Clem}. At the transition to the IMS, a kink and sudden decrease is observed, which is consistent with the changing $\mathbf{q}$ of the Bragg peaks~\cite{1977Christen}. For Nb-hp the intensity drops down to zero where the Meissner state is reached, while the other samples retain scattering down to the lowest temperature, due to the flux freezing. The transition of the homogeneous VL into the domain structure of the IMS is visible as well as a significant drop in the VL coherence $\xi_{VL}$ (panels (c, f, i), in units of $a_{VL}$). For all samples, the IMS shows a value of $\xi_{VL}\approx 3-4 \, a_{VL}$. However, the drop is most pronounced in Nb-hp, where the high quality allows $\xi_{VL}\approx 20 \,a_{VL}$ in the SS. 

Most notably, the results for $a_{VL}$ were identical for FC and FC/FH measurements in all samples as well as for ZFC measurements of Nb-hp. Additionally, several samples of Nb-mp, with different polishing / etching, yielded identical results. This independence from the measurement history seen for $a_{VL}$ is present in the integrated intensity and $\xi_{VL}$ as well, however, with minor deviations. 

% VSANS

\begin{figure}[h]
	\includegraphics[width=8cm]{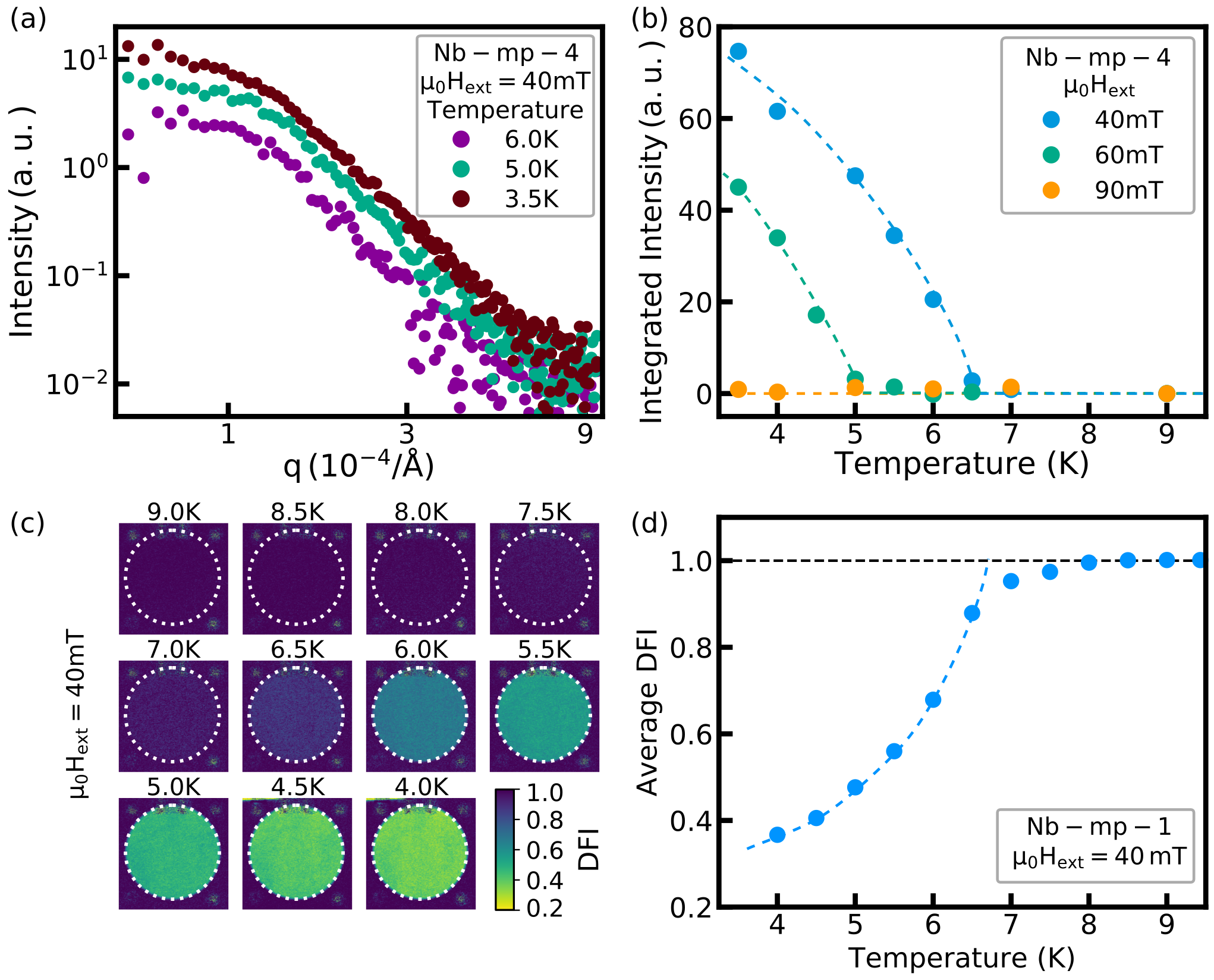}
	\caption{\label{fig:image4} VSANS and NGI. (a) $\mathbf{q}$-dependence of the VSANS intensity of Nb-mp-4 at different temperatures during FC in $40\,$mT. The data was corrected for a background at $9\,$K. (b) shows the integrated intensity for FC in three different magnetic fields. (c) FC NGI measurement series in $40\,$mT of Nb-mp-1 (indicated by a white circle). The average DFI value is plotted in (d). Colored lines are guides to the eye.}
\end{figure}

Results of VSANS and NGI measurements are presented in Fig.~\ref{fig:image4} for Nb-mp. The VSANS measurements were performed on thinner Nb-mp-4, to prevent multiple scattering, with the magnetic field applied perpendicular to the sample face. The data was corrected for a high temperature background ($T=9\,$K) and radially averaged. Outside the IMS, the data remained identical to the background. The additional signal in the IMS is attributed to scattering from the magnetic domains. Panel (a) shows data obtained in a field of $40\,$mT during FC. The scattering intensity levels off below $q\approx2\cdot 10^{-4}/\text{\AA}$, corresponding to length scales in the range above $\sim 5\,\mu$m. For a given external field, this $q$-value remains unchanged with temperature, while the intensity increases during cooldown. The temperature dependence of the integrated intensity is plotted in panel (b) for FC in three magnetic fields, where $\mu_0H_{ext}=90\,$mT lies entirely outside the IMS. \\

% NGI

NGI data from a FC measurement series of Nb-mp-1 at $\mu_0H_{ext}=40\,$mT is shown in Fig.~\ref{fig:image4}(c, d). Panel (c) contains the dark field images (DFI), normalized to high temperature data ($T=10\,$K). Below $7\,$K, scattering in the VSANS regime, i.e. from the IMS domains, causes the DFI contrast to drop below unity. The scattering is homogeneous over the whole sample and increases with decreasing temperature. The corresponding temperature dependence of the average DFI contrast is shown in panel (d), resembling the behavior of the VSANS results (panel (b)). Similar results were obtained for various fields and temperatures from Nb-mp-1, Nb-mp-2 and Nb-lp-1. In contrast, Nb-hp-1 was previously studied, showing inhomogenities connected to demagnetization effects~\cite{2015Reimann}.

%******************************************
%******************************************

% ------------------------------------------------------------------------------
% Discussion
% ------------------------------------------------------------------------------

Combining SANS, VSANS, NGI and VSM, we have studied the IMS transition from a homogeneous VL into domains with emphasis on a FC path. For a discussion, we first focus on medium purity Nb-mp. From the constant magnetization below $T_f$ strong pinning causing a freezing transition of the VL is apparent. SANS measurements show another transition at $T_{IMS}(B)<T_f(B)$, which is, however, not apparent in the magnetization data: the IMS transition is most obvious from the vortex separation $a_{VL}$, which changes from a constant value matching the frozen flux to a temperature dependent $a_{VL}^{IMS}$. $a_{VL}^{IMS}$ is independent of the external field and represents the equilibrium vortex separation due to the attractive vortex interaction. Remarkably, the pinning observed in the magnetization has no impact on the continuous vortex rearrangement in the IMS. This discrepancy requires either a breakdown of pinning, presumably by the vortex attraction in the IMS, or separate pinning effects governing the freezing and IMS transition, namely surface and bulk pinning, respectively. The vortex aggregation seen in $a_{VL}^{IMS}$ is accompanied by a breakup of the homogeneous VL into domains, as is evident in the emergent VSANS and NGI signal as well as the change of $\xi_{VL}$. The scale of the IMS domains is in the range of $5-10\mu$m, as estimated by the VSANS $q$-dependence and the NGI correlation length. Consistently, SANS indicates a VL coherence length of $\xi_{VL}\approx 0.5-2\,\mu$m in the IMS. The vortex rearrangement causes an increasing magnetic contrast of the VL domains with respect to the surrounding Meissner phase, which is reflected in the growing VSANS and NGI signal. The temperature independence of the domain size suggests that, instead of a nucleation and growth of domains, the IMS transition can be understood as a gradual phase separation. In this context, the model of spinodal decomposition \cite{1985Furukawa} has been successfully used to describe IMS domain scattering \cite{2017Reimann}. NGI measurements reveal, that the IMS transition is completely homogeneous in the whole sample, including the edges. The constant magnetization during FC, caused by flux freezing, eliminates influences of the sample geometry. As a result, the properties in the IMS are solely determined by temperature, applied field and material properties. 

In comparison, Nb-lp exhibits an equivalent behavior, where only the transition temperatures $T_c$ and $T_{IMS}$ are reduced due to lower purity. Accordingly, for Nb-hp, these transitions are increased slightly due to the higher purity. However, the exceptional quality of Nb-hp inhibits a flux freezing transition, revealing the expected diamagnetism. The associated flux expulsion reduces the vortex density which is evident in the increasing $a_{VL}$, prior to the decrease in the IMS. $a_{VL}^{IMS}$, however, is not affected by the increasing diamagnetism. The effects of flux expulsion can be observed in NGI measurements and spatially resolved SANS as well \cite{2015Reimann}. 

The universality of the IMS transition is most obvious in $a_{VL}^{IMS}$. All samples can in fact be fitted by the phenomenological expression $a_{VL}^{IMS}(t) = a_{VL}^{IMS}(t=0)\cdot (1-t^{3-t})^{1/2}$, with $t = T/T_c$, derived from numerical solutions for the superconducting penetration depth $\lambda_L$ in BCS-theory~\cite{1959Muehlschlegel}. This has recently been used to describe $a_{VL}^{IMS}$ \cite{2014Pautrat}. The connection to $a_{VL}^{IMS}$ is straightforward, as primarily $\lambda_L$ determines the vortex shape and thus its interaction potential. With decreasing sample purity and accordingly increasing $\lambda_L$, $a_{VL}^{IMS}$ is shifted to higher values which results in a lower $T_{IMS}$~\cite{1953Pippard}. For all sample purities, FC and FC/FH yielded identical results for $a_{VL}^{IMS}$ underscoring its independence of the measurement history. For Nb-Hp, this was the case even for ZFC. In contrast, in Nb-mp and Nb-lp, ZFC leads to a strongly inhomogeneous field distribution due to high pinning and sample geometry given by the critical state model~\cite{2014Gokhfeld}. Accordingly, SANS data did not show an ordered VL in this case.

% ------------------------------------------------------------------------------
% Conclusions
% ------------------------------------------------------------------------------

In conclusion, we have studied the transition from a homogeneous VL into VM domains in the IMS in Nb. This transition exhibits a universal behavior for a broad range of the sample quality and is independent of the experimental history. Driven by a vortex attraction closely related to the superconducting penetration depth, the VL contracts and breaks up into domains in a gradual phase separation, reminiscent of a spinodal decomposition scenario. The formation of an IMS, even in samples exhibiting strong pinning, suggests a breakdown of pinning due to the vortex attraction. 

\begin{acknowledgments}

We express our gratitude to S. Mayr and A. Kriele for their support with sample preparation as well as Z. Revay and C. Stieghorst for providing the sample characterizations using NAA. Further thanks are due to T. Reimann, T. Neuwirth for fruitful discussions and support with the experiments. 

\end{acknowledgments}

\bibliographystyle{unsrt}

%====================================================================

%====================================================================

%\newpage

\section*{Supplemental Material}

%******************************************
\begin{table*}
\begin{tabular}{c||cc|cccc}
Sample & $RRR$ & Impurities & Shape & Dimensions [mm] & $D$ & Experiment \\ \hline
Nb-hp-1 & $>$10k & $20\,$ppm Ta & rod & $d=5.5$ , $t=19.7$ & 0.14 ($\parallel$) & VSM \\
&&& & & 0.43 ($\perp$) & SANS \\
\hline
Nb-mp-1 & $>$300 & $150\,$ppm Ta & disc & $d=25$ , $t=1.3$ & 0.90 & SANS, NGI\\
Nb-mp-2 &&& disc & $d=10$ , $t=1.3$ & 0.79 & SANS \\
Nb-mp-3 &&& disc & $d=5$ , $t=0.3$ & 0.89 & VSM \\
Nb-mp-4 &&& strip & $20 \times 2 \times 0.2$ & 0.87 & VSANS \\
Nb-mp-5 &&& cuboid & $4.0 \times 3.8 \times 1.9$ & 0.24 & VSM \\
\hline
Nb-lp-1 & $\approx$100 & $200\,$ppm Ta & disc & $d=20$ , $t=0.6$ & 0.94 & SANS\\
Nb-lp-2 && $350\,$ppm W & cuboid & $4.0 \times 3.7 \times 1.9$ & 0.24 & VSM\\

\end{tabular}
\caption{\label{tab:samples} Single crystal niobium samples (hp: high purity, mp: medium purity, lp: low purity). All mp and lp samples were cut from the same crystal, respectively. Dimensions are given either as diameter (d) and thickness (t) for a cylindric shape or as the three edge lengths for cuboid shape. Demagnetization factors ($D$) were calculated for all discs with the magnetic field perpendicular to the surface, for the cuboids with the field parallel to the longest edge and for the rod shaped Nb-hp-1 two values are given with the field parallel ($\parallel$) and perpendicular ($\perp$) to the cylinder axis.}

\vspace{2cm}

\begin{tabular}{c||l|l|l}
Experimental & \multicolumn{1}{c|}{Instrument} & \multicolumn{1}{c|}{Evaluated} & \multicolumn{1}{c}{Measurement} \\
method & \multicolumn{1}{c|}{resolution} & \multicolumn{1}{c|}{quantities} & \multicolumn{1}{c}{details} \\ 
\hline
\begin{tabular}{c} SANS \\ ----- \\ SANS-1 \end{tabular}
& \begin{tabular}{l} $\bullet$ real space \\ \hspace{2mm} $5\,$mm \\ $\bullet$ reciprocal space \\ \hspace{2mm} $7.5\cdot10^{-4}/\text{\AA}$  \end{tabular}
& \begin{tabular}{l} VL Bragg scattering: \\ $\bullet$ vortex separation $a_{VL}$ \\ $\bullet$ scattering intensity \\ $\bullet$ VL correlation length $\xi_{VL}$  \end{tabular}
& \begin{tabular}{l}The Bragg peak positions are defined by the VL arrange- \\ ment (structure factor). The scattering intensity \\ depends on the local field distribution of the single \\ vortices (form factor), which is $q$- and temperature \\ dependent~\cite{2019Muehlbauer}. In the IMS, the intensity depends \\ further on the reduced area of SS regions, while the peaks \\ broaden due to the finite size of VL domains. Note, that \\ the observed correlation length is much larger for small \\ angle diffraction~\cite{2010Grigoriev}, compared to diffuse SANS~\cite{1998Felber}. \end{tabular} \\
\hline

\begin{tabular}{c} VSANS \\ ----- \\ KWS-3 \end{tabular}
& \begin{tabular}{l} $\bullet$ real space \\ \hspace{2mm} $10\,$mm \\ $\bullet$ reciprocal space \\ \hspace{2mm} $4\cdot10^{-5}/\text{\AA}$\end{tabular}
& \begin{tabular}{l}domain scattering function: \\ $\bullet$ domain correlation length \\ $\bullet$ domain morphology\end{tabular}
& \begin{tabular}{l}The $q$-dependent scattering function depends on the size \\ and shape of the IMS domains. The intensity is defined \\ by the contrast of the averaged magnetic flux in the SS \\ and MS regions. Contrary to the diffraction by the VL, \\ the correlation length of the VSANS measurements is \\ determined as for the diffuse scattering \\ experiments~\cite{1998Felber} \end{tabular} \\
\hline

\begin{tabular}{c} NGI \\ ----- \\ ANTARES \end{tabular}
& \begin{tabular}{l} $\bullet$ real space \\ \hspace{2mm} $0.8\,$mm \\ $\bullet$ correlation length \\ \hspace{2mm} $\xi_{NGI}=1.9\,\mu$m \end{tabular}
& \begin{tabular}{l} domain correlation function: \\ $\bullet$ real space domain \\ \hspace{2mm} distribution\end{tabular}
& \begin{tabular}{l} The DFI contrast seen with NGI depends on the IMS \\ domain correlation function at one specific distance, the \\ NGI correlation length $\xi_{NGI}$, which gives the typical \\ length scale of structures for which the NGI is most \\ sensitive to~\cite{2014Strobl}. A changing DFI can indicate a change \\ of the domain size, shape or contrast. The contrast is \\ defined by the average magnetic flux in the SS and MS \\ regions. \end{tabular}

\end{tabular}
\caption{\label{tab:techniques} Summary of the neutron based experimental methods. The list contains the resolution limits of the instruments used in this study, both in real and reciprocal space. Additionally, the quantities evaluated from the raw data for this study are listed. Furthermore, short description of the measurement techniques are given, highlighting their dependence on the properties of the vortex matter.}

\end{table*}

%******************************************

This appendix contains additional details about the experiments performed in the scope of the presented work. Table 1 lists all samples that were used, including information about their purity, shape, size, demagnetization factor and experimental methods they were used for.
As the superconducting properties of niobium are essentially isotropic, in particular for medium purity samples~\cite{1972Almond, 1975Novotny, 1977Christen}, the crystal orientation of the individual samples do not have an impact on the performed experiments. However, for completeness the orientations are as follows:
Nb-hp-1 has a [110] direction along the cylinder axis. The beam was aligned along a [110] direction perpendicular to the cylinder axis. For Nb-mp, all disc-shaped samples have a cylinder axis close to [100] (off by $2-3^{\circ}$). For Nb-lp, all disc-shaped samples have a cylinder axis along [110]. The orientation of the cuboids used for magnetization measurements (Nb-mp-5 and Nb-lp-2) as well as the strip used for VSANS (Nb-mp-4) have not been determined. 

Table 2 summarizes the neutron based measurement techniques, listing the relevant resolution limits in real and reciprocal space, as well as the magnetic properties they are sensitive to and the physical quantities they yield. 

The VL-correlation length $\xi_{VL}$ obtained from SANS measurements was determined from the width of the first order Bragg-peaks $\Delta q_{Bragg}$ in $\mathbf{q}$-direction. It was corrected by the reciprocal space resolution $\Delta q_{Inst}$ according to~\cite{1995Yaron, 2010Grigoriev} as $\Delta q = \sqrt{\Delta q_{Bragg}^2 - \Delta q_{Inst}^2}$. The VL-correlation length is then defined as $\xi_{VL} = (2\pi ) / (\Delta q)$

\end{document}